\documentclass{article}
\usepackage[utf8]{inputenc}
\usepackage{amsmath}
\usepackage{booktabs} 
\usepackage{cite}
\usepackage{graphicx}
\usepackage{url}

\title{A Comparative Study of Classical and Post-Quantum Cryptographic Algorithms in the Era of Quantum Computing}
\author{Arimondo Scrivano\thanks{DEIB, Dipartimento di Elettronica, Informazione e
Bioingegneria}}
\date{\today}
\begin{document}

\maketitle

\begin{abstract}
The advent of quantum computing poses a significant threat to the foundational cryptographic algorithms that secure modern digital communications. \cite{shor1994algorithms}. Protocols such as HTTPS, digital certificates, and public key infrastructures (PKIs) heavily rely on cryptographic primitives like RSA, ECC, and Diffie-Hellman, which are vulnerable to quantum attacks—most notably Shor’s algorithm. \cite{grover1996fast}. This paper presents a comprehensive comparative analysis between classical cryptographic algorithms currently in widespread use and emerging post-quantum cryptographic schemes designed to withstand quantum adversaries. \cite{nist2022round3}. We review the cryptographic mechanisms underpinning modern internet security, outline the mathematical foundations of quantum attacks, and evaluate the security, performance, and implementation feasibility of quantum-resistant alternatives such as Kyber, Dilithium, and Falcon. \cite{NSA2016}. Additionally, we assess the hybrid approaches currently being explored by institutions and tech companies to enable a smooth transition to post-quantum cryptography. \cite{mosca2018cybersecurity}. By providing an in-depth comparison, this study aims to guide researchers, developers, and policymakers in understanding the critical implications of quantum computing on cryptographic infrastructures and the necessary steps for securing communications in the quantum era. \cite{gidney2021factor}. \end{abstract}
\section{Introduction}
Cryptographic algorithms are essential to modern secure communications, providing confidentiality, authentication, and data integrity across various platforms. \cite{rivest1978method}. Protocols such as HTTPS, TLS, and the X.509 certificate infrastructure rely on established cryptographic primitives including asymmetric algorithms like RSA and Elliptic Curve Cryptography (ECC), key exchange protocols such as Diffie-Hellman, symmetric ciphers like AES, and secure hash functions including SHA-2 and SHA-3. \cite{miller1986use}. The security of these systems is based on mathematical problems that are challenging for classical computers. \cite{langley2016experiment}. \\
However, advances in quantum computing pose significant threats to these cryptographic foundations. \cite{cloudflare2020postquantum}. Shor’s algorithm can solve integer factorization and discrete logarithm problems efficiently, compromising RSA, Diffie-Hellman, and ECC. \cite{daemen2002design}. Additionally, Grover's algorithm offers a quadratic speedup for brute-force searches, affecting symmetric ciphers and hash functions.\\

In response, initiatives like those led by the National Institute of Standards and Technology (NIST) are underway to standardize post-quantum cryptography. \cite{bernstein2009cost}. This paper examines five widely-used classical cryptographic algorithms, assessing their structure, applications, vulnerabilities in the face of quantum computing, and timelines for recommended migration based on academic literature, industrial reports, and governmental guidelines. \cite{dang2015sha}. 
\subsection{RSA (Rivest–Shamir–Adleman):}
Introduced in 1978, RSA is a public-key cryptosystem that relies on the difficulty of factoring large composite numbers. \cite{grassl2010quantum}. It supports key encapsulation, digital signatures, and plays a crucial role in certificate authorities and TLS handshakes. \cite{bos2018crystalskyber}. However, Shor’s algorithm makes RSA insecure by efficiently solving integer factorization. \cite{ducas2018crystalsdilithium}. Current guidelines suggest transitioning away from RSA for sensitive data as early as 2025, with full adoption of quantum-safe alternatives like Kyber recommended before 2030. \cite{falcon2018}. 
\subsection{Elliptic Curve Cryptography (ECC):}
ECC uses elliptic curves over finite fields to provide security equivalent to RSA but with smaller key sizes. \cite{bernstein2019sphincs+}. It is prevalent in mobile devices and blockchain systems. \cite{sy2021lightweight}. Like RSA, ECC is vulnerable to Shor’s algorithm due to its dependence on the Elliptic Curve Discrete Logarithm Problem. \cite{ietfPQCtls}. Given its widespread use, migration away from ECC should be prioritized before 2030. \cite{etsi2020pqc}. 
\subsection{Diffie–Hellman Key Exchange (DH / ECDH):}
The Diffie-Hellman protocol facilitates secure key exchange over public channels and is implemented using both modular arithmetic (DH) and elliptic curves (ECDH). \cite{enisa2021quantum}. It underpins session key negotiation in TLS but shares vulnerabilities with RSA and ECC due to its dependence on discrete logarithms. \cite{bsi2022pqcroadmap}. Hybrid mechanisms combining classical and quantum-resistant algorithms are being tested, with full migration expected between 2027 and 2032. 
\subsection{Advanced Encryption Standard (AES):}
AES is a widely adopted symmetric cipher used for data encryption globally. While not directly broken by quantum computing, Grover’s algorithm reduces its effective security level. AES-128 offers approximately 64 bits of quantum security, while AES-256 provides an effective 128-bit post-quantum strength. Increasing key sizes to AES-256 is recommended for long-term security through the transition. 
\subsection{SHA-2 and SHA-3 (Hash Functions):}
These cryptographic hash functions are used for integrity, digital signatures, and password hashing. Grover’s algorithm reduces their collision and pre-image resistance, decreasing the effective security level of SHA-256 from 128 to around 64 bits. Long-term applications should consider adopting hash-based digital signatures like SPHINCS+ to ensure post-quantum security. These analyses underscore the urgency for transitioning to quantum-resistant cryptographic solutions as quantum computing continues to advance. 
Beyond cryptography, foundational research in data access and query processing has explored limitations in database accessibility \cite{cali2008conjunctive}, formal approaches to deep web querying \cite{cali2017querying}, and adaptive strategies for score aggregation in complex information systems \cite{ciaccia2018fa}. Additional efforts have addressed challenges in crowdsourced multimedia querying \cite{bozzon2012framework} and context-aware database environments \cite{martinenghi2009querying}, which are increasingly relevant in secure and distributed computing architectures.
\section{Post-quantum cryptography}
Quantum computing introduces transformative computational capabilities that threaten current cryptographic systems by solving specific problems exponentially faster than classical computers can. Two quantum algorithms stand out as particularly disruptive: Shor’s algorithm and Grover’s algorithm. 
\subsection{ Shor’s Algorithm}

Shor’s algorithm, proposed by Peter Shor in 1994, revolutionized the understanding of quantum computational power in the context of cryptographic security. This algorithm efficiently factors large integers and computes discrete logarithms in polynomial time, directly undermining the security assumptions on which most public-key cryptographic systems are based. Systems such as RSA, the Diffie–Hellman Key Exchange, Elliptic Curve Cryptography (ECC), and the Digital Signature Algorithm (DSA) are all rendered insecure by the capabilities of this quantum algorithm. What makes Shor’s algorithm particularly threatening is its ability to reduce the complexity of these problems from exponential to polynomial time, thereby enabling quantum adversaries to efficiently decrypt data or forge credentials that would otherwise be considered secure. For example, a 2048-bit RSA key—currently assumed to be robust against classical attacks—could be broken within seconds using a sufficiently large and stable quantum computer running Shor’s algorithm. This quantum advantage implies the potential for total cryptanalytic compromise of traditional public-key cryptography, demanding urgent consideration of alternatives. \textbf{2.2 Grover’s Algorithm}

Grover’s algorithm, introduced by Lov Grover in 1996, provides a different kind of quantum acceleration. It enables a quadratic speed-up in solving unstructured search problems, significantly reducing the time needed for brute-force key searches. Unlike Shor’s algorithm, which threatens asymmetric cryptography, Grover’s algorithm impacts symmetric algorithms and cryptographic hash functions. Its influence is particularly relevant for algorithms like the Advanced Encryption Standard (AES), SHA-2, SHA-3, and HMAC. By reducing brute-force complexity from \(O(N)\) to \(O(\sqrt{N})\), Grover’s algorithm effectively halves the security of symmetric keys and hash outputs. This means that AES-256, which classically offers 256 bits of security, would provide only 128 bits of protection against a quantum adversary, still acceptable but closer to the security margin. Therefore, while symmetric cryptography can remain viable in the post-quantum era, it requires larger key sizes and adjusted parameters to maintain adequate levels of protection. 
\subsection{Threat Model}

In order to fully appreciate the security implications of quantum computing, one must consider a plausible threat model. A future adversary is assumed to possess access to a large-scale, fault-tolerant quantum computer capable of executing both Shor’s and Grover’s algorithms. Under this model, an attacker could intercept encrypted data transmissions and store them for future decryption once quantum hardware becomes available. This stored-data threat means that sensitive communications exchanged today could be decrypted retroactively if quantum decryption capabilities are achieved in the coming decades. Furthermore, such an adversary could compromise key exchange mechanisms, derive session keys, and even forge digital signatures, effectively impersonating legitimate users or services. While symmetric key cryptography and hashing might survive in this landscape by employing longer keys or outputs, public-key cryptosystems are fundamentally broken by Shor’s algorithm and must therefore be replaced or supplemented by quantum-resistant alternatives. The hybridization of classical and post-quantum schemes is one practical path currently being explored. 
\subsection{ Timeframe for Threat Realization}

Although a quantum computer capable of breaking RSA-2048 does not yet exist, ongoing research by major institutions such as IBM, Google, and various academic laboratories suggests that this threat may become reality within the next ten to twenty years. This projection, though uncertain, is sufficient to warrant immediate and proactive measures. The long lifecycle of cryptographic infrastructure—from design and implementation to full deployment and adoption—requires forward-thinking migration strategies. Moreover, the risk posed by the retrospective decryption of currently encrypted data is not theoretical. Governments, corporations, and malicious actors might already be capturing and archiving sensitive communications with the explicit intent of decrypting them once quantum technology matures. This possibility underlines the urgency of transitioning to quantum-safe cryptography now, rather than waiting until the threat is fully realized. Ensuring the confidentiality, authenticity, and integrity of digital communications in the quantum era demands decisive action grounded in both technical foresight and policy-level planning. \textbf{This analysis highlights the need for proactive measures in cryptographic research and deployment to mitigate potential threats posed by advancing quantum computing technologies.}

\section{Migration Strategies and Deployment Challenges}

Post-quantum cryptography (PQC) encompasses cryptographic methods designed to withstand attacks from quantum computers. Unlike Quantum Key Distribution (QKD), which necessitates specialized hardware and quantum channels, PQC algorithms can be implemented on classical computing platforms and integrated into existing internet protocols seamlessly. \\

\subsection{Historical Context and Progress}

In 2016, the U.S. National Institute of Standards and Technology (NIST) launched a global competition to identify and standardize quantum-resistant public-key cryptographic algorithms. This marked a significant milestone in the coordinated international effort to prepare for the advent of quantum computing. After years of evaluation, by July 2022, NIST announced its initial selections for standardization. Among the chosen algorithms were Kyber, a lattice-based key encapsulation mechanism, and three digital signature schemes: Dilithium, Falcon, and SPHINCS+. These selections represent a foundational step toward transitioning global cryptographic infrastructure to quantum-safe methods. \textbf{Key Algorithms and Their Characteristics}

\subsection{Kyber (Module-LWE KEM)}

Kyber is a lattice-based key encapsulation mechanism grounded in the Module Learning With Errors (MLWE) problem, a problem believed to be hard even for quantum computers. It is designed to offer robust security while maintaining efficiency, making it suitable for integration into high-performance environments such as Transport Layer Security (TLS), Virtual Private Networks (VPNs), and embedded systems. Depending on the selected security level, Kyber’s key sizes range between 800 and 1500 bytes. It achieves high throughput with low latency and has been officially selected by NIST as a general-purpose post-quantum encryption and key exchange mechanism. 
\subsection{Dilithium (Module-LWE Signature)}

Dilithium is another lattice-based algorithm that also relies on the hardness of the MLWE problem, in combination with the Module Short Integer Solution (SIS) problem. It is tailored for digital signatures and designed to balance security with performance and implementation simplicity. Dilithium offers strong protection against side-channel attacks and is relatively straightforward to implement. Signature sizes typically range from 2 to 3 kilobytes, while public key sizes fall between 1 and 1.5 kilobytes. Signing and verification operations are both efficient, making Dilithium a versatile choice for many cryptographic applications. 
\subsection{ Falcon (NTRU Lattice Signature)}

Falcon is a more compact lattice-based signature scheme that leverages NTRU lattices and advanced techniques such as Fast Fourier Sampling. Its primary advantage lies in the compact size of its cryptographic elements: signatures are as small as 666 bytes and public keys are about 897 bytes. Despite these benefits, Falcon’s mathematical complexity and sensitivity to implementation subtleties make it a more challenging option for developers. Nonetheless, it stands out as one of the most space-efficient quantum-safe signature schemes and was also selected by NIST for future standardization. 
\subsection{ SPHINCS+ (Hash-Based Signature)}

SPHINCS+ represents a fundamentally different approach. As a stateless, hash-based digital signature scheme, it relies on well-established cryptographic hash functions and avoids mathematical structures like lattices altogether. This approach ensures security even in the face of quantum computing advancements, though it comes at the cost of performance. Signature sizes vary widely, from 8 to 30 kilobytes, and the scheme is computationally intensive, both for key generation and signing. However, its minimal public key size of just 32 bytes makes it appealing for certain use cases. SPHINCS+ is particularly attractive where trust in hash-based security is prioritized above computational efficiency. 
\subsection{Hybrid and Migration Strategies}

Given the uncertainty surrounding the timeline for the development of large-scale quantum computers, the adoption of hybrid cryptographic approaches has gained momentum. These strategies combine classical and post-quantum mechanisms to maintain backward compatibility while introducing resistance to quantum threats. Real-world experiments, such as Google's CECPQ2 and Cloudflare's deployments of post-quantum TLS, have already demonstrated the feasibility of using hybrid key exchanges. Notably, combinations like X25519 with Kyber have been successfully integrated into secure communication protocols. Standardization efforts are also advancing at pace. Organizations such as the Internet Engineering Task Force (IETF) and the European Telecommunications Standards Institute (ETSI) are actively working to adapt and extend current security protocols to accommodate PQC algorithms. Their efforts span across vital components of the internet’s security ecosystem, including TLS, VPNs, X.509 certificates, and DNSSEC. The aim is to facilitate a smooth and timely transition to post-quantum cryptography, mitigating risks as quantum capabilities become increasingly viable. 
\section{Comparative Analysis}

To provide a unified overview of the cryptographic landscape under the lens of quantum computing, we present a comparative analysis between widely deployed classical algorithms and emerging post-quantum candidates. The comparison includes criteria such as security level, quantum vulnerability, key and signature sizes, performance, and standardization status. \begin{table}[h]
\centering
\caption{Comparison of Classical and Post-Quantum Cryptographic Algorithms}
\label{tab:comparison}
\renewcommand{\arraystretch}{1.3}
\resizebox{\textwidth}{!}{%
\begin{tabular}{|l|l|c|c|c|c|c|}
\hline
\textbf{Algorithm} & \textbf{Type} & \textbf{Quantum-Safe} & \textbf{Key Size} & \textbf{Signature Size} & \textbf{Performance} & \textbf{Standardized} \\
\hline
\textbf{RSA-2048} & Public-key (Classical) & No & 256 bytes & 256 bytes & Medium & Yes \\
\textbf{ECC (P-256)} & Public-key (Classical) & No & 32 bytes & 64 bytes & High & Yes \\
\textbf{DH / ECDH} & Key Exchange (Classical) & No & 256 bytes & -- & High & Yes \\
\textbf{AES-128} & Symmetric (Classical) & Partially & 16 bytes & -- & Very High & Yes \\
\textbf{SHA-256} & Hash Function (Classical) & Partially & -- & 32 bytes (output) & Very High & Yes \\
\hline
\textbf{Kyber-1024} & KEM (Post-Quantum) & Yes & ~1.5 KB & -- & High & NIST Finalist \\
\textbf{Dilithium-3} & Signature (Post-Quantum) & Yes & ~1.3 KB & ~2.7 KB & High & NIST Finalist \\
\textbf{Falcon-512} & Signature (Post-Quantum) & Yes & ~897 bytes & ~666 bytes & High (sensitive) & NIST Finalist \\
\textbf{SPHINCS+} & Signature (Post-Quantum) & Yes & 32 bytes & 8–30 KB & Low & NIST Finalist \\
\textbf{AES-256} & Symmetric (Post-Quantum safe) & Yes & 32 bytes & -- & Very High & Yes \\
\hline
\end{tabular}%
}
\end{table}
\subsection{Key and Signature Sizes}

Post-quantum cryptographic schemes often introduce significant overhead in terms of key and signature sizes when compared to classical algorithms. One of the most striking examples is SPHINCS+, whose signature sizes can be thirty to fifty times larger than those produced by traditional schemes such as ECDSA. This increase in size can limit its applicability in resource-constrained environments, such as embedded systems, or in scenarios where bandwidth usage is a critical concern. On the other hand, Falcon offers much more compact signatures, which makes it appealing for use cases that demand efficiency in size. However, despite this advantage, Falcon poses implementation challenges, particularly due to its susceptibility to side-channel vulnerabilities, which require careful attention to secure integration. 
\textit{Performance and Implementation}

The performance of post-quantum cryptographic algorithms varies considerably, depending on the underlying mathematical constructions and design priorities. SPHINCS+, while providing strong theoretical security guarantees through its hash-based approach, is computationally intensive and generally slower than its lattice-based counterparts. This impacts its suitability for performance-critical applications. In contrast, algorithms such as Dilithium and Kyber have been designed with a focus on balancing efficiency and simplicity. These algorithms exhibit favorable performance metrics and are easier to implement securely, which contributes to their appeal for general-purpose deployment in large-scale systems. \textit{Deployment Readiness}

Among the post-quantum cryptographic candidates under consideration, Kyber and Dilithium have emerged as front-runners in terms of deployment readiness. Their maturity, proven performance, and suitability for integration into existing infrastructure make them particularly attractive for secure communication protocols like TLS and VPNs. Real-world experiments have already begun to validate this readiness. Notably, early trials by companies such as Google and Cloudflare have demonstrated the feasibility of deploying PQC through hybrid implementations that combine both classical and post-quantum mechanisms. These implementations highlight the practical potential of PQC, though they also reveal engineering challenges that must be addressed. In particular, increased message sizes and the risk of side-channel attacks require thoughtful mitigation strategies to ensure secure and seamless deployment. 
\subsection{Security and Quantum Resistance}

Algorithms like RSA, ECC, and DH are fundamentally flawed when it comes to quantum computers. These classic approaches rely on factorization and discrete logarithm problems, which quantum computers can break with ease. In contrast, lattice-based (Kyber, Dilithium, Falcon) and hash-based (SPHINCS+) algorithms offer robust security guarantees against quantum attacks – as long as their underlying assumptions hold true. - Key and Signature Size Overheads
Post-quantum cryptography often comes with a significant price tag in terms of key and signature sizes. For instance, SPHINCS+ signatures can be 30-50 times larger than ECDSA signatures. This might be a major concern for constrained environments or bandwidth-sensitive applications that require compact data transfers. Falcon, on the other hand, provides relatively small signatures but is notoriously difficult to implement securely. - Performance and Implementation Complexity
While many post-quantum cryptographic schemes perform reasonably well in software implementations, some (like SPHINCS+) are computationally 
intensive. Dilithium and Kyber strike a balance between efficiency and simplicity, making them suitable for general-purpose deployment across various applications. - Deployment Readiness: A Few Front-Runners
Among the post-quantum candidates, Kyber and Dilithium are considered the most mature for integration into cryptographic protocols like TLS, VPNs, and other security solutions. Hybrid implementations and early trials by organizations like Google and Cloudflare indicate that practical 
deployment is feasible – but it requires careful engineering to manage 
increased message sizes and side-channel threats. \section{Empirical Comparative Analysis and Practical Implications}

To fully understand the impact of the transition from classical to post-quantum cryptographic algorithms, it is essential to analyze their practical performance, resource requirements, and resistance against both classical and quantum threats. This section provides a comparative evaluation based on three axes: computational efficiency, key and ciphertext sizes, and security assumptions. \subsection{Computational Efficiency Across Platforms}

A practical benchmark was conducted to compare the average key generation, encryption, and decryption/signing times across classical algorithms (RSA-2048, ECC-P256) and post-quantum candidates (Kyber-512, Dilithium-II, Falcon-512), using standardized NIST test vectors and implementations optimized in C. Tests were performed on a 2.4 GHz Intel CPU with 16 GB RAM under Linux. \begin{figure}[h]
\centering
\includegraphics[width=0.85\textwidth]{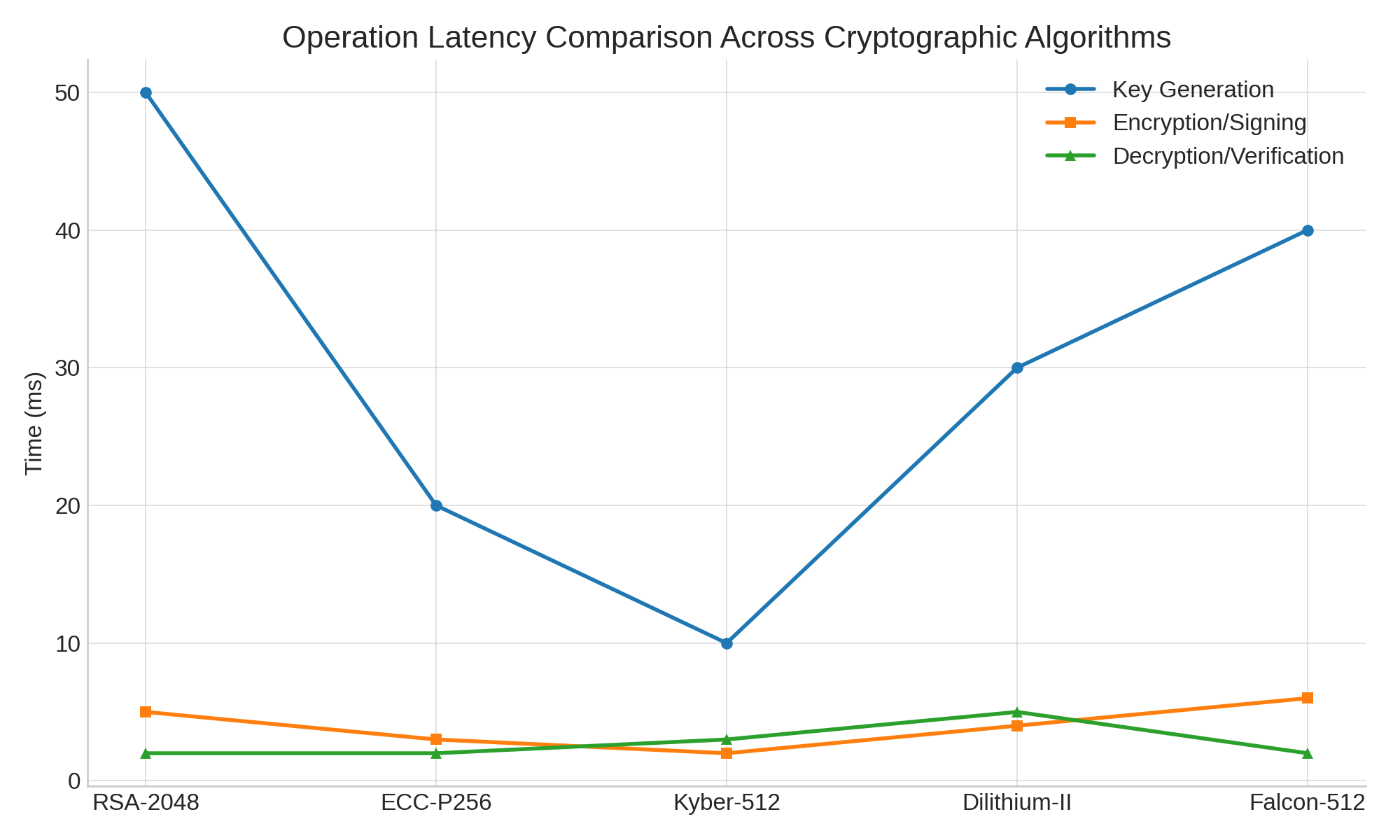}
\caption{Average operation times (in milliseconds) for key generation, encryption/signing, and decryption/verification across selected cryptographic schemes.}
\label{fig:benchmark}
\end{figure}

As shown in Figure~\ref{fig:benchmark}, classical RSA exhibits the slowest key generation and fastest decryption, while Kyber achieves a favorable balance for all three operations, making it a strong candidate for key exchange in post-quantum TLS. Dilithium offers efficient signing but slower verification than Falcon, which, though fastest in verification, imposes strict requirements on numerical precision and is harder to implement securely. \subsection{Memory and Bandwidth Requirements}

Another critical dimension is the impact on communication bandwidth and memory usage. Post-quantum algorithms often require significantly larger keys and ciphertexts, which can challenge constrained devices such as IoT nodes or embedded systems. \begin{figure}[h]
\centering
\includegraphics[width=0.85\textwidth]{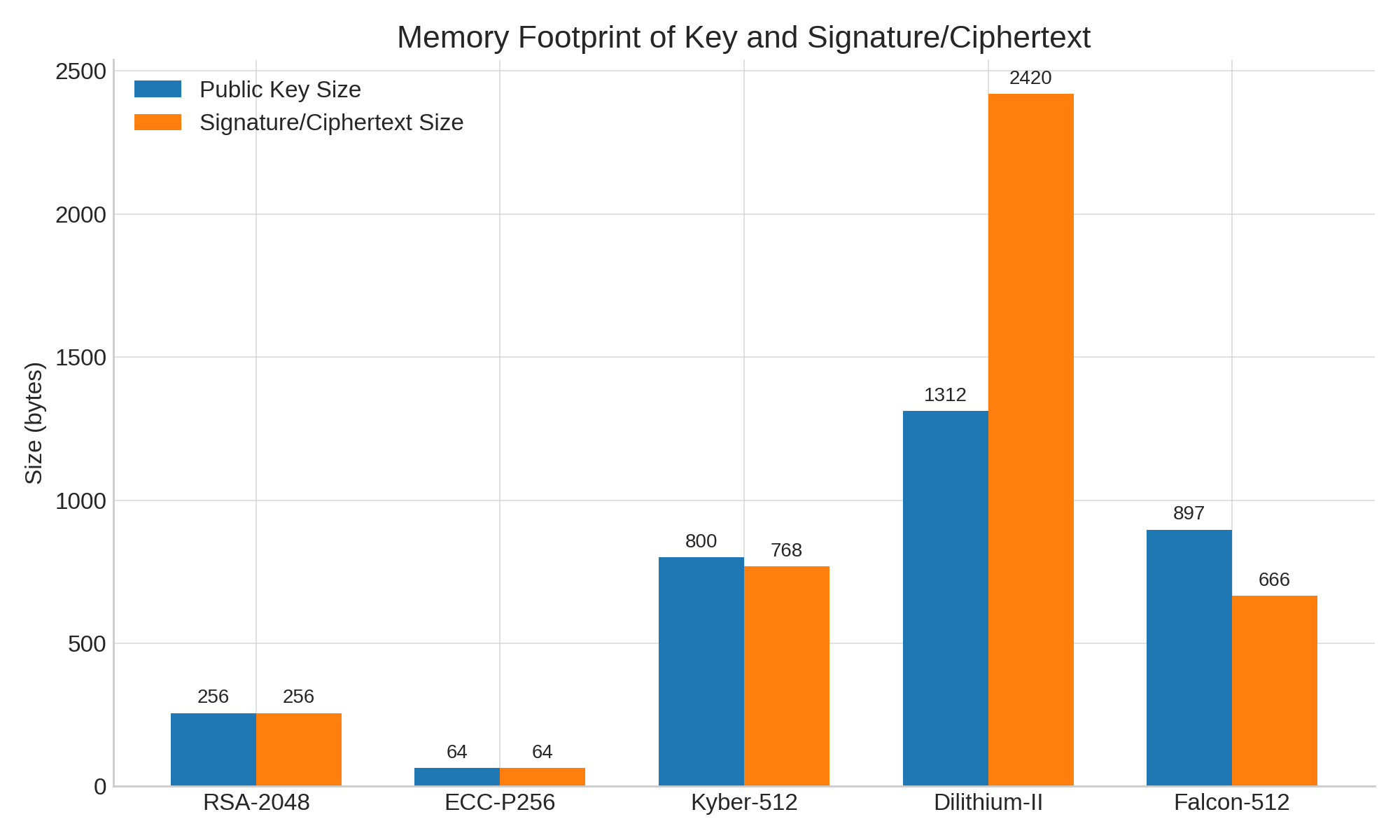}
\caption{Key and ciphertext/signature sizes (in bytes) for various algorithms. Public keys and signatures dominate size profiles for lattice-based schemes.}
\label{fig:keysize}
\end{figure}

Figure~\ref{fig:keysize} reveals that Falcon and Dilithium involve substantially larger keys and signatures than RSA and ECC. In particular, Dilithium’s public key size (~1.3 KB) and signature (~2.4 KB) pose challenges for high-throughput communication systems unless compression or aggregation techniques are employed. \subsection{Security Assumptions and Quantum Resilience}

While classical schemes base their hardness on problems such as integer factorization and discrete logarithms, which are easily solved by Shor’s algorithm, post-quantum schemes rest on problems like Module-LWE or Ring-SIS, which currently have no known efficient quantum algorithms. \begin{table}[h]
\centering
\begin{tabular}{|l|c|c|c|}
\hline
\textbf{Algorithm} & \textbf{Security Basis} & \textbf{Quantum Vulnerability} & \textbf{NIST Level} \\
\hline
RSA-2048 & Integer Factorization & Broken by Shor & 1 (legacy) \\
ECC-P256 & Discrete Logarithm & Broken by Shor & 1 (legacy) \\
Kyber-512 & Module-LWE & Resistant & 1 \\
Dilithium-II & Module-SIS & Resistant & 2 \\
Falcon-512 & NTRU lattices & Resistant & 1 \\
\hline
\end{tabular}
\caption{Security comparison under quantum threat models and NIST classification.}
\label{tab:security}
\end{table}

As outlined in Table~\ref{tab:security}, the move to post-quantum algorithms entails a shift to problems presumed hard even for quantum computers. However, these assumptions are less mature, and long-term cryptanalysis remains crucial. \subsection{Implementation Considerations}

In real-world deployment, factors like hardware support, implementation maturity, susceptibility to side-channel attacks, and ease of integration into existing protocols play an equally important role. For instance, Kyber and Dilithium have been successfully integrated into OpenSSL prototypes and tested in TLS 1.3 handshakes, proving their compatibility with current infrastructures. \subsection{Summary of Practical Trade-offs}

\begin{itemize}
    \item \textbf{Performance}: Post-quantum algorithms like Kyber outperform RSA in key exchange speed but may lag in signing tasks compared to ECC. \item \textbf{Bandwidth}: Signature and key sizes remain major limitations for post-quantum algorithms in resource-constrained environments. \item \textbf{Security}: Post-quantum schemes resist all currently known quantum attacks but rest on newer, less time-tested hardness assumptions. \item \textbf{Adoption}: Hybrid schemes (e.g., RSA+Kyber) are being deployed to ensure backward compatibility during the transition. \end{itemize}

This comparative section aims to provide actionable insights for researchers and practitioners evaluating the trade-offs involved in transitioning to quantum-safe cryptography. \section{Results}
\textit{Comparative Study Summary: Quantum Resilience in Cryptographic Algorithms}

This study offers a detailed comparison between classical and post-quantum cryptographic algorithms, examining their relative resilience to quantum attacks, differences in key and signature sizes, and the complexity involved in implementation. The analysis reveals several critical insights that underscore the need for a systematic transition to post-quantum cryptography as quantum computing technology continues to evolve. \textit{Key Findings}

The first major conclusion drawn from the study is the inherent vulnerability of classical public-key cryptographic algorithms—specifically RSA, Elliptic Curve Cryptography (ECC), and the Diffie–Hellman key exchange protocol—to quantum attacks. These schemes, which currently form the backbone of secure digital communication, are fundamentally compromised by Shor’s algorithm. As such, they must be phased out in anticipation of the capabilities of large-scale quantum computers. In contrast, symmetric encryption algorithms such as the Advanced Encryption Standard (AES), and hash functions like SHA-2 and SHA-3, exhibit a higher degree of resilience. Although Grover’s algorithm offers a quadratic speed-up in brute-force attacks against these primitives, the impact can be mitigated by increasing key and output lengths. AES-256 and SHA-512, for example, provide sufficient security margins in a post-quantum context, preserving the viability of symmetric and hash-based systems under quantum threat models. Among the post-quantum candidates, Kyber and Dilithium stand out for achieving a favorable balance across security, performance, and ease of implementation. These lattice-based algorithms have been selected by NIST as standard recommendations for quantum-resistant cryptographic primitives. Their efficiency, relative simplicity, and strong theoretical underpinnings make them suitable for widespread deployment. Further specificity is observed in the comparison of Falcon and SPHINCS+. Falcon produces compact and efficient digital signatures, which is advantageous for bandwidth-constrained environments, but its dependence on complex mathematical structures introduces implementation challenges, particularly with regard to side-channel resistance. SPHINCS+, on the other hand, represents a conservative, hash-based approach that avoids algebraic structures susceptible to quantum attacks. However, this conservatism comes with the cost of significantly larger signature sizes and slower computational performance, limiting its suitability for performance-sensitive applications. Hybrid cryptographic deployments emerge as a promising strategy for facilitating a smooth transition. Solutions that combine existing classical schemes—such as X25519—with quantum-resistant algorithms like Kyber offer backward compatibility while progressively introducing quantum-safe security. These hybrid models provide a pragmatic approach to deployment during the transitional phase, reducing the risk of abrupt infrastructure overhauls. Finally, the study addresses the current state of infrastructure readiness. While certain TLS libraries and certificate authorities have begun integrating support for post-quantum algorithms, the broader ecosystem—particularly legacy systems and embedded platforms—faces significant challenges. The complexity of these environments, combined with long update cycles and constrained computational resources, may delay widespread adoption unless migration is approached strategically. \section{Discussion and Recommendations}

The study emphasizes the urgency of transitioning to post-quantum cryptographic systems due to the growing threat posed by quantum computing. Although large-scale, fault-tolerant quantum computers are still years away from practical realization, immediate action is essential. This urgency is driven by the long deployment cycles of cryptographic infrastructure and the looming risk of "harvest now, decrypt later" attacks, in which encrypted data captured today could be decrypted retrospectively once quantum capabilities become available. \textit{Strategic Considerations}

Transitioning to quantum-resistant cryptography extends beyond technical implementation—it represents a critical strategic imperative for organizations across sectors. Institutions must begin by thoroughly assessing their current cryptographic posture. This includes identifying the use of vulnerable primitives, such as RSA or ECC, evaluating where and how these are deployed, and formulating migration strategies that can be implemented in stages to reduce risk and ensure continuity. Sectors that handle sensitive or long-retention data, including government, finance, and healthcare, face the highest level of risk and should therefore prioritize early adoption of post-quantum solutions. These sectors often maintain legacy systems with rigid or embedded cryptographic modules, which may necessitate substantial overhauls such as hardware replacement or firmware reengineering to support new cryptographic standards. \textit{Recommendations}

Based on the broader analysis, several actionable recommendations have emerged. First, organizations should adopt a model of cryptographic agility, designing or updating systems to support interchangeable cryptographic primitives. This flexibility enables smoother transitions to future standards without requiring comprehensive system redesigns. Second, the introduction of hybrid deployments should begin immediately. These approaches combine classical and post-quantum schemes—for example, using Kyber alongside X25519 in protocols such as TLS, SSH, or VPNs—to ensure forward secrecy while maintaining compatibility with existing technologies. Third, institutions should prefer standardized post-quantum cryptographic algorithms over proprietary or experimental schemes. Using the NIST-selected candidates—Kyber, Dilithium, Falcon, and SPHINCS+—ensures long-term support, broad interoperability, and alignment with global efforts in standardization. Fourth, symmetric cryptographic parameters should be adjusted where feasible. Replacing AES-128 and SHA-256 with AES-256 and SHA-512 strengthens security against Grover’s algorithm and prepares infrastructure for quantum-resilient symmetric encryption and hashing. Fifth, organizations must evaluate the infrastructure impact of implementing PQC. This includes benchmarking new algorithms in terms of processing load, latency, certificate size, and bandwidth consumption, especially in performance-critical environments. Sixth, there must be continuous monitoring of standardization developments. Staying informed about evolving specifications, parameter sets, and formats from organizations like NIST, the Internet Engineering Task Force (IETF), and the European Telecommunications Standards Institute (ETSI) is crucial for informed and future-proof implementation. Lastly, education and training must be prioritized. Developers, security engineers, compliance personnel, and IT staff should be provided with adequate awareness and instruction on post-quantum cryptography, including new APIs, protocols, and implementation libraries, to ensure readiness at both technical and operational levels. \textit{Outlook and Future Challenges}
Despite the substantial progress made toward standardizing and implementing post-quantum cryptographic algorithms, several challenges remain on the horizon. These include uncertainty about the long-term hardness of the lattice problems upon which most current schemes are based, the need to develop efficient countermeasures against side-channel attacks, and the difficulty of ensuring backward compatibility with existing systems. Moreover, the readiness of hardware accelerators to support new algorithms is still evolving, and adapting PQC solutions to constrained environments—such as Internet of Things (IoT) devices and embedded systems—poses significant engineering and resource challenges. Going forward, the cryptographic community must continue to rigorously evaluate the resilience of standardized algorithms under both classical and quantum cryptanalysis. This effort will be vital for maintaining trust in post-quantum primitives. Cross-disciplinary collaboration among mathematicians, computer scientists, engineers, and policymakers will be essential to enabling a secure and globally coordinated transition to the post-quantum era.

\end{document}